\documentclass[review]{elsarticle}

\usepackage{amsfonts}
\usepackage{amsmath}
\usepackage{graphicx,setspace,algorithmic,algorithm,multirow,cases,amssymb,subfigure,setspace}
\usepackage{lettrine}
\usepackage{multirow}
\usepackage{xcolor}
\usepackage{times}
\usepackage{graphicx}
\usepackage{amsmath}
\usepackage{indentfirst}
\usepackage{booktabs}
\usepackage{blindtext}
\usepackage{array}
\usepackage{tabularx,booktabs}
\usepackage{color, soul}
\usepackage{bm}
\usepackage{array}
\usepackage{booktabs}

\journal{Journal of Neurocomputing}

\begin{document}

\begin{frontmatter}
\title{Audio Description from Image by Modal Translation Network}

\author[a,b]{Hailong Ning}
\author[a]{Xiangtao Zheng\corref{d}}
\cortext[d]{Corresponding author}
\author[c]{Yuan Yuan}
\author[a]{Xiaoqiang Lu}

\address[a]{Key Laboratory of Spectral Imaging Technology CAS, Xi'an Institute of Optics and Precision Mechanics, Chinese Academy of Sciences, Xi'an 710119, Shaanxi, P. R. China.}
\address[b]{University of Chinese Academy of Sciences, Beijing 100049, P. R. China.}
\address[c]{The Center for OPTical IMagery Analysis and Learning (OPTIMAL), School of the Computer Science, Northwestern Polytechnical University, Xi'an 710072, Shaanxi, P. R. China.}

\begin{abstract}
Audio is the main form for the visually impaired to obtain information. In reality, all kinds of visual data always exist, but audio data does not exist in many cases. In order to help the visually impaired people to better perceive the information around them, an image-to-audio-description (I2AD) task is proposed to generate audio descriptions from images in this paper. To complete this totally new task, a modal translation network (MT-Net) from visual to auditory sense is proposed. The proposed MT-Net includes three progressive sub-networks: 1) feature learning, 2) cross-modal mapping, and 3) audio generation. First, the feature learning sub-network aims to learn semantic features from image and audio, including image feature learning and audio feature learning. Second, the cross-modal mapping sub-network transforms the image feature into a cross-modal representation with the same semantic concept as the audio feature. In this way, the correlation of inter-modal data is effectively mined for easing the heterogeneous gap between image and audio. Finally, the audio generation sub-network is designed to generate the audio waveform from the cross-modal representation. The generated audio waveform is interpolated to obtain the corresponding audio file according to the sample frequency. Being the first attempt to explore the I2AD task, three large-scale datasets with plenty of manual audio descriptions are built. Experiments on the datasets verify the feasibility of generating intelligible audio from an image directly and the effectiveness of proposed method.
\end{abstract}


\end{frontmatter}


\section{Introduction}
\begin{figure}[tp]
\begin{center}
\includegraphics[width=0.7\linewidth]{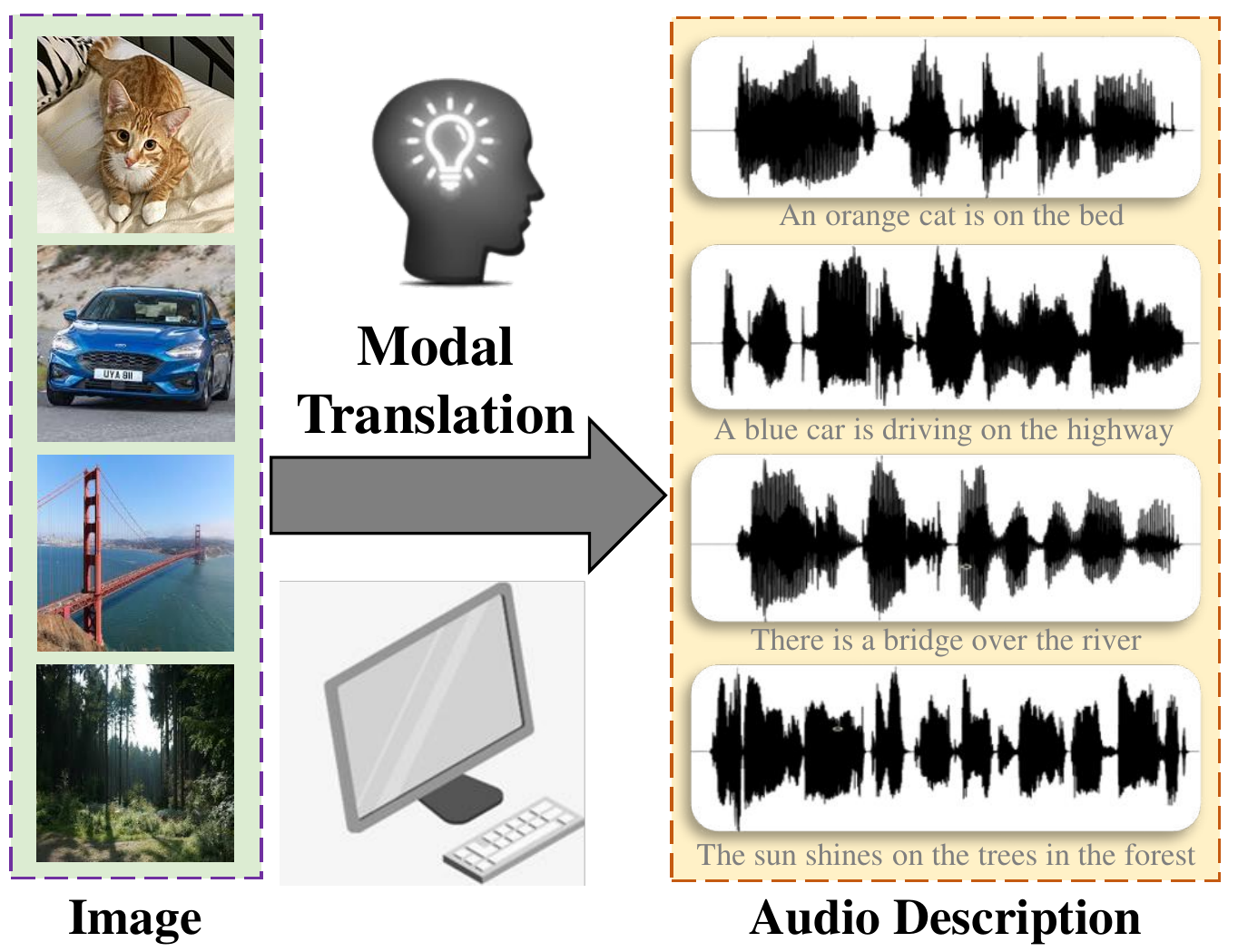}
\renewcommand{\figurename}{Fig.}
\end{center}
    \caption{\small{The schematic diagram of the image-to-audio-description task, which aims to generate audio descriptions (right) from image (left).}}
\label{fig:1}
\end{figure}

There are millions of visually impaired people all over the world, and how to assist them to better perceive the environment is a significant but challenging task. In the past decades, extensive works have been made in sensory substitution by the neurocognitionists \cite{chebat2018sensory, brown2016audio, striem2012reading}. Specifically, they attempt to train visually impaired people activating their visual cortex by auditory messages based on the cross-modal plasticity \cite{striem2012reading, glick2017cross}. However, it is difficult and needs a lot of time to train the visually impaired people, which is unrealistic at least for now. With the rapid development of computer technology and machine learning \cite{ijcai_LiCNW17, ijon_ZhengZL20, aaai_LiCNW17}, especially deep learning \cite{zheng2020joint, zhang2020feature, lu2020sound}, an alternative thought through sensory translation is expected to help the visually impaired people better perceive the environment. Based on this thought, an image-to-audio-description (I2AD) task is proposed to generate an audio description from a visual image (see Figure \ref{fig:1}) in this paper.

The I2AD task aims to generate an audio description from a visual image, which could be fundamental for many practical applications \cite{owens2016ambient, caraiman2017computer, alamri2019audio}. In virtual reality, audio descriptions can be generated for virtual scenes automatically to enhance the experience of immersion \cite{8683318}. For people who are blind, have low vision, or who are otherwise visually impaired, the generated audio descriptions can involve the accessibility of the visual images \cite{walczak2017creative, walczak2018audio}. In the field of human-computer interaction, the I2AD task can provide the opportunity for human to communicate with machines in the audio fashion \cite{zhao2017multimodal}. In addition, the I2AD task may be applied in many other cases potentially \cite{8269806}, making the society more convenient and intelligent.

The I2AD task belongs to the category of cross-modal generation, which is more challenging than the current hot cross-modal retrieval task. Concretely, the cross-modal retrieval task aims to explore the joint representations of multi-modal data \cite{suris2018cross, wang2017adversarial, shang2019adversarial}, which only needs to retrieve samples that exist in a database. In contrast, the cross-modal generation task tries to generate a new data of one modality when given a data sample characterized by one or more other modalities \cite{chen2017deep, wan2019towards, zhou2019talking}. In other word, the cross-modal generation task requires learning a complex generative function that produces meaningful outputs. In addition, the I2AD task is also more challenging than the existing vision-to-audio works, such as generating the audio of different instruments \cite{hao2018cmcgan, chen2017deep}, generating ambient audios from visual data \cite{zhou2018visual}, and generating percussion of people hitting objects with a drumstick \cite{chen2018visually, owens2016visually}. These existing vision-to-audio works do not need to consider the intelligibility of the generated audio, while the generated audio description by the I2AD task must be intelligible natural language. That is to say, I2AD task is not only to understand the content of the input image, but also to translate the information contained in the image as intelligible natural language in audio form (audio description). As a result, the main challenges for the I2AD task are the heterogeneous gap between two modalities and the generation of intelligible natural language in audio form.

To explore the I2AD task and overcome the two main challenges mentioned above, a modal translation network (MT-Net) from visual to auditory sense is proposed by exploring the inherent relationship of image and audio. Considering that the raw inputs are with enormous variability (such as illumination variation in visual inputs and different emotional state in audio inputs), the image information are translated into the audio information in the feature level rather than the raw data level. In this way, the heterogeneous gap can be mitigated effectively. The MT-Net includes three main progressive sub-networks: 1) feature learning, 2) cross-modal mapping, and 3) audio generation. First, the feature learning sub-network aims to learn semantic features from image and audio, including image feature learning and audio feature learning. Second, the cross-modal mapping sub-network transforms the image feature into a cross-modal representation with the same semantic concept as the audio feature. In this way, the correlation between image and audio is effectively mined for easing the heterogeneous semantic gap. Finally, the audio generation sub-network is designed to generate the audio waveform from the cross-modal representation. The generated audio waveform can be interpolated to obtain the corresponding intelligible natural language in audio form according to the sample frequency. It should be noted that as the first attempt, this paper only explores the simple form in the I2AD task to generate single words in audio form.

To sum up, the main contributions of this paper are threefolds:
\begin{itemize}
  \item An image-to-audio-description (I2AD) task is proposed to generate audio descriptions from images, which is fundamental for many applications. To explore the I2AD task, three large-scale audio caption datasets are built.
  \item A modal translation network (MT-Net) from visual to auditory sense is proposed for the I2AD task by exploring the inherent relationship of image and audio. The MT-Net generates the audio description by translating the image information to the audio information in the feature level.
  \item An audio generation sub-network is designed to generate intelligible and natural audio description. The audio generation sub-network adopts the 1D convolution kernel with holes to model the complex relationship of different phonemes in the audio.
\end{itemize}

The remaining parts of this paper are organized as follows: Section \ref{relatedworks} reviews the related works. Section \ref{TheProposedmethod} gives a detailed description of the proposed network. The experiments are shown in Section \ref{experiments}. Finally, a conclusion is presented in Section \ref{conclutions}.

\section{Related works}\label{relatedworks}

This work closely relates to the existing researches in cross-modal learning. Cross-modal learning aims to learn the relationship between different modalities. Significant progress has been observed in visual, audio, and language modality learning, including cross-modal retrieval \cite{deng2018triplet, guo2019jointly, xu2019deep}, cross-modal matching \cite{harwath2018jointly, nagrani2018seeing}, image captioning \cite{zhu2018image, xiao2019daa, zhang2020image}, visual question answering \cite{gordon2018iqa, chao2018cross, ruwa2019mood}, video summarization \cite{9091191, cvpr_ZhaoLL18, zhong2019video}, {\it etc.} This paper focuses on the cross-modal learning between audio and visual modalities. According to the research task, the cross-modal learning between audio and visual modalities can be divided: 1) audio-visual correspondence, 2) audio source localization and separation in visual scenes, 3) audio-to-vision, and 4) vision-to-audio.
\begin{figure*}[tp]
\begin{center}
\includegraphics[width=\linewidth]{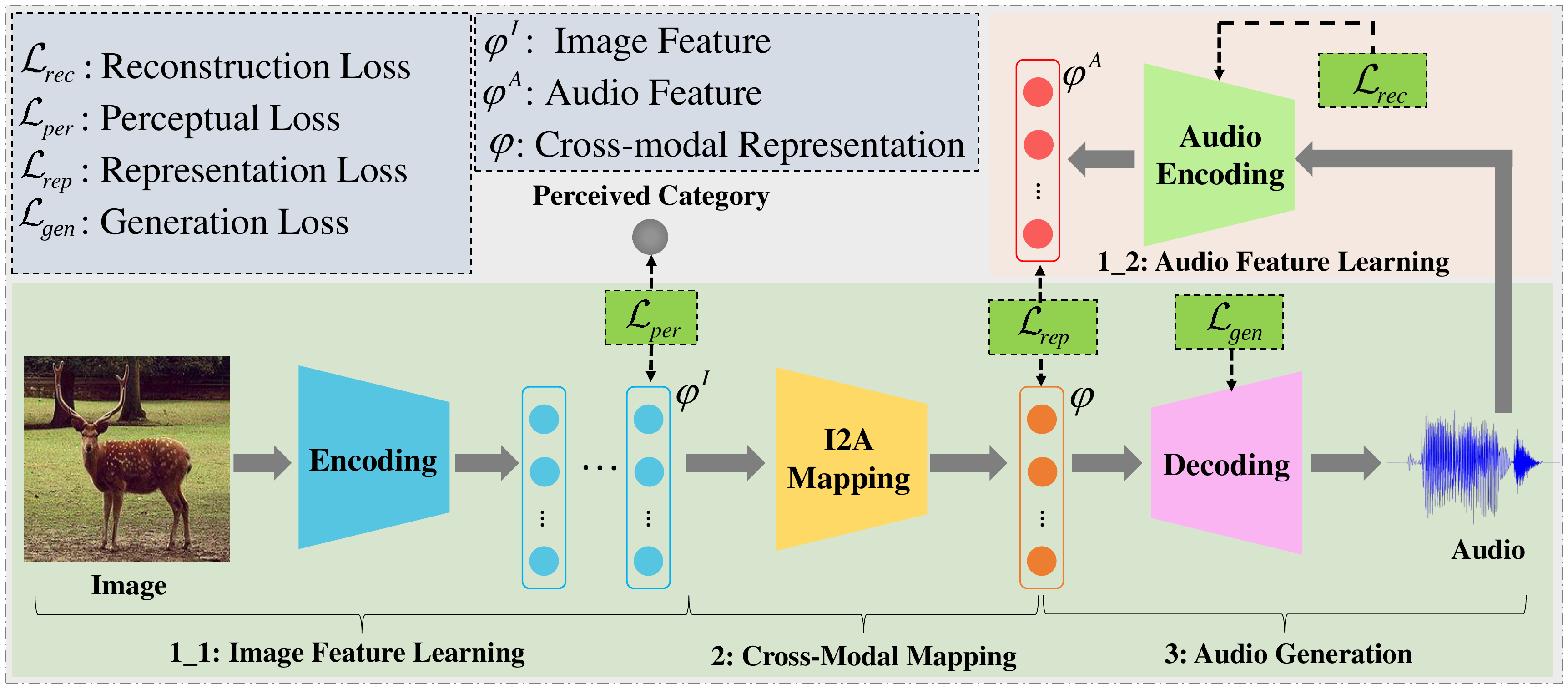}
\renewcommand{\figurename}{Fig.}
\end{center}
    \caption{\small{The proposed network. First, a feature learning sub-network is built to learn semantic features from image and audio, including image feature learning and audio feature learning. Second, a cross-modal mapping sub-network transforms the image semantic feature into a cross-modal representation with the same semantic concept as the audio semantic feature. Finally, an audio generation sub-network is designed for generating the audio waveform from the cross-modal representation.}}
\label{fig:2}
\end{figure*}

\noindent
{\bf Audio-visual correspondence} is the most widely studied problem in the cross-modal learning about audio and visual modalities. Arandjelovi\'{c} and Zisserman \cite{arandjelovic2017look} introduce an audio-visual correspondence learning task without any additional supervision. Harwath {\it et al.} \cite{harwath2018jointly} explore a neural network model that learns to associate segments of spoken audio captions with the semantically relevant portions of natural images. Nagrani {\it et al.} \cite{nagrani2018seeing} introduce a cross-modal biometric matching task to select the face image corresponding to the voice or determine the corresponding voice related to the face image.

\noindent
{\bf Audio source localization and separation in visual scenes} includes two sub-tasks. The audio source localization sub-task is to reveal the audio source location in visual scenes \cite{Senocak_2018_CVPR}. The audio separation sub-task is to separate the mixed audio signals of several objects in visual scenes \cite{zhao2018sound}. Arandjelovi\'{c} and Zisserman \cite{arandjelovic2018objects} design a network for localizing the audio source in an image by embedding the input audio and image into a common space. Zhao {\it et al.} \cite{zhao2018sound} introduce a PixelPlayer to locate image regions which produce audios by leveraging large amounts of unlabeled videos. In addition, the PixelPlayer is also able to separate the input audios into a set of components that represents the audio from each pixel.

\noindent
{\bf Audio-to-vision} aims to generate images related to the input audio. Ginosar {\it et al.} \cite{ginosar2019learning} present a cross-modal translation network from monologue speech to corresponding conversational gesture motion. Wan {\it et al.} \cite{wan2019towards} propose a conditional generative adversarial network (GAN) to generate images from audios. The network is able to adjust the output scale according to the volumes changes of the input audio. Oh {\it et al.} \cite{oh2019speech2face} put forward a Speech2Face network to infer a person's appearance from a short audio recording of that person speaking.

\noindent
{\bf Vision-to-audio} generates the audio corresponding to visual input. Chen {\it et al.} \cite{chen2017deep} design two separate conditional GANs for audio spectrograms generation from instrument's images and instrument's images generation from audio spectrograms, respectively. Subsequently, Hao {\it et al.} \cite{hao2018cmcgan} build a CMCGAN network using cycle GAN for  audio spectrograms generation from instrument's images and instrument's images generation from audio spectrograms. The CMCGAN unifies the vision-to-audio and audio-to-vision tasks into a common framework and improves the output quality. Zhou {\it et al.} \cite{zhou2018visual} present a network for generating ambient audios from given input video frames. The generated audios are fairly realistic and have good temporal synchronization with the visual inputs. Owens {\it et al.} \cite{owens2016visually} propose a network for producing percussion of people hitting objects with a drumstick. Chen {\it et al.} \cite{chen2018visually} explore to generate fine-grained audio from a variety of audio classes. To improve the quality of generated audio, Chen {\it et al.} leverage pre-trained audio classification networks for aligning the semantic information between the generated audio and its ground truth. Another strategy to generate audio descriptions from images is image-to-text and then text-to-audio. The strategy firstly generates the text description based on the image caption model \cite{iccv_LiuT0G19, ijon_DingQXW20}, and then transforms the text description into corresponding audio based on some existing TTS software. However, the generated audio description based on this strategy can not express some implicit information, such as emotion.

This paper focuses on generating descriptive audios similar to human speech from given images. Note that the proposed I2AD task is more challenging than the existing vision-to-audio works since the existing vision-to-audio works do not need to consider the intelligibility of the generated audio.

\section{The proposed network}\label{TheProposedmethod}
In this work, a modal translation network (MT-Net) from visual to auditory sense is proposed for the I2AD task to generate audio descriptio from images. Since the raw inputs are with enormous variability, the image information are translated into the audio information in the feature level instead of the raw data level. As is depicted in Figure \ref{fig:2}, the proposed MT-Net adopts an encoder-decoder architecture, including three main progressive sub-networks: 1) feature learning, 2) cross-modal mapping, and 3) audio generation. First, the feature learning sub-network learns image semantic feature $\varphi^{I}$ and audio semantic feature $\varphi^{A}$ from two modalities. Here, a 2D convolution kernel (see Figure \ref{fig:3}(a)) with holes is adopted for capturing more context information  so as to learn more discriminant feature. Second, the cross-modal mapping sub-network transforms the image semantic feature $\varphi^{I}$ to a cross-modal representation $\varphi$ with the same semantic concept as the audio semantic feature $\varphi^{A}$. Finally, the audio generation sub-network is designed to generate the audio waveform ${\bf{a}}^{gen}$ from the cross-modal representation $\varphi$. To obtain high-quality audio waveform, a 1D convolution kernel with holes (see Figure \ref{fig:3}(b)) is adopted for capturing long range correlations between distant phonemes in audios. The generated audio waveform is interpolated to obtain the corresponding audio file according to the sample frequency.

\subsection{Feature learning}\label{FeatureLearning}
This sub-network is composed of image feature learning and audio feature learning parts. The image feature learning part aims to encode the input image into a low-dimension semantic feature, and the audio feature learning part is designed to learn a condensed intermediate feature of the audio. The details about the image feature learning and audio feature learning are introduced as follows.
\subsubsection{\textbf{Image feature learning}}\label{ImageEncoder}
\begin{figure}[tp]
\begin{center}
\includegraphics[width=0.5\linewidth]{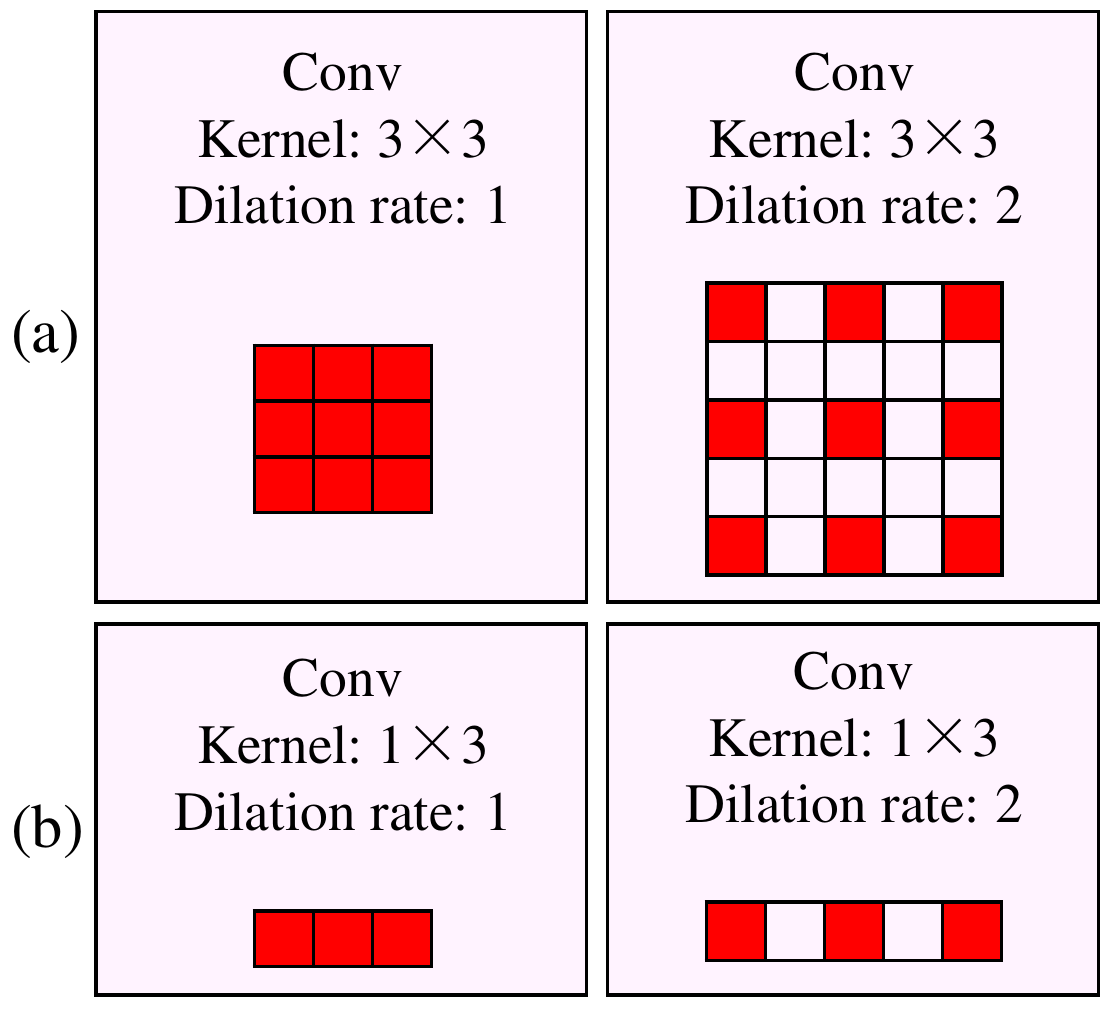}
\renewcommand{\figurename}{Fig.}
\end{center}
    \caption{\small{Convolution kernel with holes. These kernels enable the layer to have a greater receptive field without increasing the memory footprint. (a) is a 2D convolution kernel, and (b) is a 1D convolution kernel.}}
\label{fig:3}
\end{figure}
The image feature learning part aims to learn the image semantic feature $\varphi^{I}$. First, following the previous works \cite{tip_JiCLJXHF20, 9091241}, the input training image sample ${\bf I}^{tr}$ is preprocessed as a standard normalization ${\bf I}_{norm}^{tr}$. Second, the standardized training image sample ${\bf I}_{norm}^{tr}$ is input into five stacked residual blocks with two fully connected (FC) layers to obtain the initial image semantic feature. Third, the initial image semantic feature is fed into a softmax layer to infer the category $\widehat{{\bf{y}}}\in \mathbb{R}^{h}$ of the input training image sample. Finally, the inferred category $\widehat{{\bf{y}}}$ is used to calculate the loss function for optimize the network to learn the final image semantic feature $\varphi^{I}$. The details of the image feature learning part is shown in Table \ref{tab:1}. The specific implementation process about the image feature learning is as follows.

First, in order to make convergence faster while training the network, the input training image sample ${\bf I}^{tr}$ is preprocessed to a standard normalization by:
 \begin{equation}\label{Eq.:1}
\begin{split}
{\bf I}_{norm}^{tr}=\frac{{\bf I}^{tr}-\mu}{max(\theta,\frac{1}{\sqrt{P}})},
\end{split}
\end{equation}
 where ${\bf I}_{norm}^{tr}$ denotes the result after standard normalization. $\mu$, $\theta$ and $P$ indicate the mean value, standard variance and pixel number of each input training image sample ${\bf I}^{tr}$, respectively.

Second, the standardized training image sample ${\bf I}_{norm}^{tr}$ flows into five stacked residual blocks with two fully connected (FC) layers where the image semantic feature $\varphi^{I}$ extraction occurs. Before data flowing into each residual block, it needs to be processed by a transition layer to adjust the channel number of feature map. The transition layer used in this work consists of a $1\times1$ convolutional layer with batch normalization operation. Let ${\bf X}_{k}^{ts}$ and  ${\bf{X}}_{k-1}^{ts}$ denote the output and input of the $k$-th transition layer, respectively. Notably, ${\bf{X}}_{k-1}^{ts}={\bf I}_{norm}^{ts}$ if $k=1$. As a result, the output of  the $k$-th transition layer can be formulated as:
 \begin{equation}\label{Eq.:2}
\begin{split}
{\bf X}_{k}^{ts}=trans\left({\bf X}_{k-1}^{ts}\right)=\mathcal{R}\left(\mathcal{BN}\left({\bf  X}_{k-1}^{ts}\ast {\bf W}_{k-1}^{ts}+{\bf b}_{k-1}^{ts}\right)\right),
\end{split}
\end{equation}
where $trans$ represents the entire operation process of the transition layer, $\mathcal{R}$ is the activation function adopted as the {\it Rectified Linear Unit} (ReLU), $\mathcal{BN}$ refers to batch normalization, and $\ast$ indicates convolution operation. ${\bf W}_{k-1}^{ts}$ and ${\bf b}_{k-1}^{ts}$ signify the convolution kernel with size $1\times1$ and bias vector in the $k$-th transition layer, which are to-be-learned parameters. Similar to the building block of ResNet \cite{he2016deep}, each residual block also adopts two coupled convolutional layer with kernel size $3\times3$ to learn a residual mapping. Differently, we introduce holes of size 2 (see Figure \ref{fig:3}(a)) in the convolution kernel to increase the receptive field for capturing more context information \cite{yu2017dilated}. Let ${\bf X}_{k}^{res}$ and ${\bf X}_{k-1}^{res}$ be the output and input of the $k$-th residual block, respectively.
Notably, ${\bf X}_{k-1}^{res}={\bf X}_{k}^{ts}$.
The operation by each residual block can be defined as:
 \begin{equation}\label{Eq.:3}
\begin{split}
{\bf X}_{k}^{res}&=res \left({\bf X}_{k-1}^{res}\right)=\mathcal{R}\left(\mathcal{F}\left({\bf X}_{k-1}^{res},{\bf W}^{res}\right)+{\bf X}_{k-1}^{res}\right)\\
&=\mathcal{R}\left(\mathcal{F}\left({\bf X}_{k}^{ts},{\bf W}^{res}\right)+{\bf X}_{k}^{ts}\right),
\end{split}
\end{equation}
where $res$ represents the entire operation process of each residual block. $\mathcal{F}(\cdot,\cdot)$, consisting of two coupled convolutional layer with kernel size $3\times3$ and hole size 2, means the residual mapping to be learned. ${\bf W}^{res}$ denotes the to-be-learned parameter. Here, the bias vector is absorbed into ${\bf W}^{res}$ for simplicity. Then, the top feature of the stacked residual blocks is squeezed in global average pooling manner \cite{cui2017kernel}. The squeezed feature is fed into two FC layers to extract the initial image semantic feature.

Third, the initial image semantic feature is fed into a softmax layer to infer the category $\widehat{{\bf{y}}}$ of the input training image sample.

 Finally, the inferred category $\widehat{{\bf{y}}}$ is used to calculate the loss function for optimize the network to learn the final image semantic feature $\varphi^{I}$. To this end, the loss function (namely perceptual loss) is adopted as follows:
 \begin{equation}\label{Eq.:4}
\begin{split}
\mathcal{L}_{per}=-\frac{1}{N}\sum_{i=1}^{N}\left({\bf{y}}_{i}^{\mathrm{T}}log(\widehat{{\bf{y}}_{i}})\right)+\lambda_{1}\sum_{l=1}^{L_{1}}\sum_{t=1}^{T_{1}}({\rm W}_{ls}^{I})^{2},
\end{split}
\end{equation}
where ${\bf{y}}\in \mathbb{R}^{h}$ denotes the human perceived category. Subscript $i$ indexes the $i$-th training data. $\lambda_{1}$ is the coefficient of weight penalty which controls the relative importance of the two terms in Eq. \ref{Eq.:4}, and ${\rm W}_{lt}^{I}$ means the value in row $l$ and column $t$ in matrix ${\bf{W}}^{I}$, which denotes all the to-be-learned parameters in the image feature learning part. $L_{1}$ and $T_{1}$ are the size dimensions of ${\bf{W}}^{I}$.

By minimizing the loss in Eq. \ref{Eq.:4}, the image feature learning part is trained. After the part is trained, the softmax layer is removed. The output of the last FC layer (namely $\rm FC_{2}^{1}$ in Table \ref{tab:1}) is extracted as the image semantic feature $\varphi^{I}$.

\subsubsection{\textbf{Audio feature learning}}
The audio feature learning aims to learn the audio semantic feature $\varphi^{A}$ for better translating the image information to the audio information in the feature level. As is shown in Figure \ref{fig:2}, first, the training audio sample ${\bf{a}}^{tr}$ is input into the multi-layer autoencoder to achieve the latent-space representation (namely the initial audio semantic feature). Second, the initial audio semantic feature is used to reconstruct the input training audio sample. Finally, the reconstructed training audio sample $\widehat{{\bf{a}}}^{tr}$ are used to calculate the loss function for optimize the network to learn the final audio semantic feature $\varphi^{A}$. The details of the audio feature learning part is summarized in Table \ref{tab:2}. The specific implementation process about the audio feature learning is as follows.

First, the training audio sample ${\bf{a}}^{tr}$ is input into the multi-layer autoencoder to obtain the initial audio semantic feature. Let $\bm{F}_{m}({\bf{a}}^{tr})$ and $\bm{F}_{m-1}({\bf{a}}^{tr})$ be the output and input of the $m$-th layer, respectively. Notably, $\bm{F}_{m-1}({\bf{a}}^{tr})={\bf{a}}^{tr}$ if $m$=1.
Thus, $\bm{F}_{m}({\bf{a}}^{tr})$ can be computed by:
\begin{equation}\label{Eq.:5}
\begin{split}
\bm{F}_{m}({\bf{a}}^{tr})=\hbar({\bf{W}}_{m-1}\bm{F}_{m-1}({\bf{a}}^{tr})+{\bf{b}}_{m-1}),
\end{split}
\end{equation}
where ${\bf{W}}_{m-1}$ and ${\bf{b}}_{m-1}$ are the to-be-learned projection matrix and bias vector. $\hbar(\cdot)$ is a non-linear activation function which is set to $tanh$ in the implementation.

Second, the initial audio semantic feature is utilized to reconstruct the input training audio sample. The reconstructed training audio sample is denoted as $\widehat{{\bf{a}}}^{tr}$.

Finally, the loss function (namely reconstruction loss) is formulated with MSE loss and weight decay term as follows:
\begin{equation}\label{Eq.:6}
\begin{split}
\mathcal{L}_{rec}=\frac{1}{2N}\sum_{i=1}^{N}\left({\bf{a}}_{i}^{tr}-\widehat{{\bf{a}}_{i}}^{tr}\right)^{2}+\lambda_{2}\sum_{l=1}^{L_{2}}\sum_{t=1}^{T_{2}}({\rm W}_{lt}^{A})^{2},
\end{split}
\end{equation}
where subscript $i$ indexes the $i$-th training data. $\lambda_{2}$ is the coefficient of weight penalty which controls the relative importance of the two terms in Eq. \ref{Eq.:6}. ${\rm W}_{lt}^{A}$ indicates the value in row $l$ and column $t$ in matrix ${\bf{W}}^{A}$, which denotes all the to-be-learned parameters in audio feature learning. $L_{2}$ and $T_{2}$ are the size dimensions of ${\bf{W}}^{A}$.

By minimizing the reconstruction loss in Eq. \ref{Eq.:6}, the audio feature learning part is trained. After the part is trained, a low-dimension audio semantic feature $\varphi^{A}$ (the output of the $\rm FC_{3}^{2}$ in Table \ref{tab:2}) is obtained for translating the information from visual modality to audio modality in the feature level.

\subsection{Cross-modal mapping} \label{CrossModal}
The cross-modal mapping sub-network aims to translate the image semantic information to audio semantic information in the feature level. In this way, the heterogeneous semantic gap between image and audio can be mitigated effectively, because the information translation in the feature level avoids the affect of the enormous variability in the raw data level. The cross-modal mapping sub-network is composed of two coupled FC layers to transfer information from visual modality to audio modality. The specific implementation process about the cross-modal mapping is as follows.

\begin{figure}[tp]
\begin{center}
\includegraphics[width=1.0\linewidth]{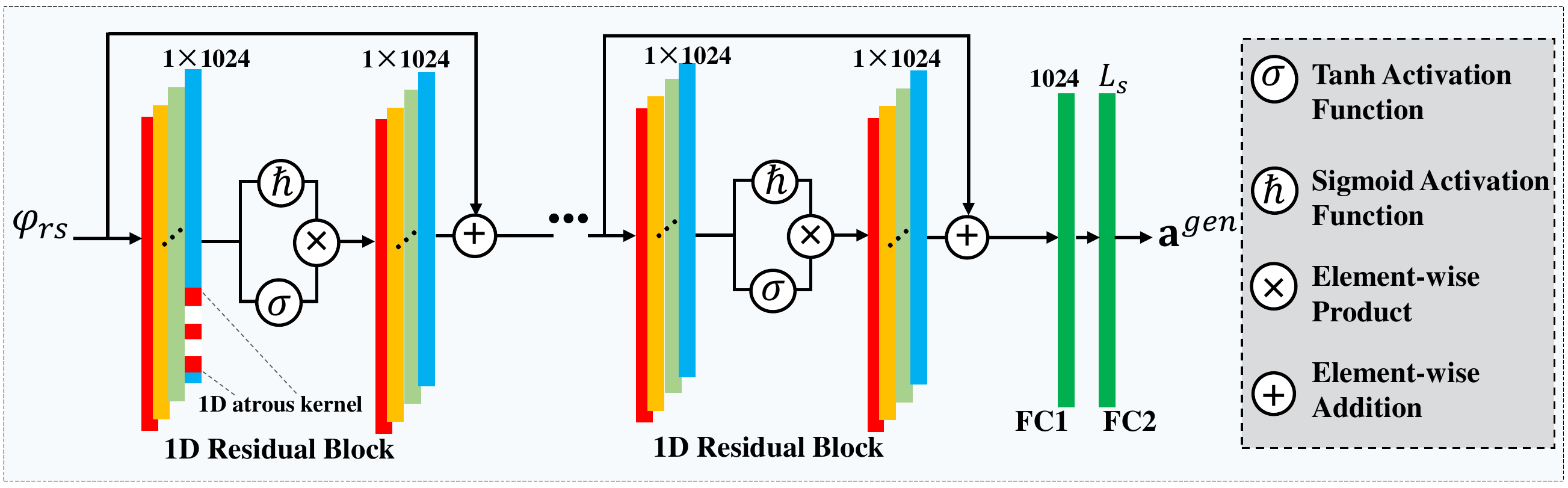}
\renewcommand{\figurename}{Fig.}
\end{center}
    \caption{\small{The designed audio generation sub-network. First, the reshaped cross-modal representation $\varphi_{rs}$ is fed into the three stacked 1D residual blocks to model the complex relationship of different phonemes. Second, the output from the last 1D residual block is sent into two additional FC layers for generating the final audio waveform ${\bf{a}}^{gen}$.}}
\label{fig:2-1}
\end{figure}

The learned image semantic feature $\varphi^{I}$ is fed into the cross-modal mapping sub-network for generating the cross-modal representation $\varphi=f(\varphi^{I})$, where is the cross-modal mapping function.
Then, the generated cross-modal representation $\varphi$ is used to calculate the loss function for optimize the cross-modal mapping sub-network. The loss function (namely representation loss) is adopted as follows \cite{chen2018visually}:
 \begin{equation}\label{Eq.:7}
\begin{split}
\mathcal{L}_{rep}=&\left(1-\sum_{i=1}^{N}\varphi^{A}_{i}\varphi_{i}/\sqrt{\sum_{i=1}^{N}(\varphi^{A}_{i})^{2}}\sqrt{\sum_{i=1}^{N}(\varphi_{i})^{2}}\right)\\
&+\lambda_{3}\sum_{l=1}^{L_{3}}\sum_{t=1}^{T_{3}}({\rm W}_{lt}^{rep})^{2},
\end{split}
\end{equation}
where subscript $i$ indexes the $i$-th training data. $\lambda_{3}$ is the coefficient of weight penalty which controls the relative importance of the two terms in Eq. \ref{Eq.:7}. ${\rm W}_{lt}^{rep}$ means the value in row $l$ and column $t$ in matrix ${\bf{W}}^{rep}$, which denotes all the to-be-learned parameters in the cross-modal mapping sub-network. $L_{3}$ and $T_{3}$ are the size dimensions of ${\bf{W}}^{rep}$.

By minimizing the loss in Eq. \ref{Eq.:7}, the input image semantic feature is transformed to a cross-modal representation with the same semantic concept as the audio feature. In this way, the correlation of inter-modal data can be mined to ease the heterogeneous semantic gap problem.

\subsection{Audio generation}
The audio generation is designed to transform the obtained cross-modal representation $\varphi$ to the corresponding audio sample. Intuitively, the audio sample is an organic combination of multiple consecutive phonemes. To model the complex relationship of different phonemes, an audio generation sub-network is devised for generating high quality audio waveform based on the cross-modal representation $\varphi$. Specifically, the audio generation sub-network adopts three stacked 1D residual blocks with 1D dilation convolution kernel and gated activation unit as the main component (see Figure \ref{fig:2-1}). The 1D dilation convolution kernel is with shape $1\times3$ and hole size 2 so as to increase the receptive field and capture long range correlations between distant phonemes (see Figure \ref{fig:3}(b)). The gated activation unit, inspired by PixelRNN \cite{van2016pixel}, is helpful to model more complex interactions for obtaining more powerful representation and generating higher quality audio waveform. The specific implementation process about the audio generation is as follows.

First, the cross-modal representation $\varphi$ is reshaped to match the input shape of the 1D convolution. The reshaped cross-modal representation is denoted as $\varphi_{rs}$.

Second, the reshaped cross-modal representation $\varphi_{rs}$ is fed into the three stacked 1D residual blocks to model the complex relationship of different phonemes. Let ${\bf E}_{n}^{tr}$ and ${\bf E}_{n-1}^{tr}$ be the output and input of the $n$-th 1D residual block, respectively.
Notably, ${\bf E}_{n-1}^{tr}=\varphi_{rs}$ if $n=1$.
The operation process of $n$-th 1D residual block can be represented by:
 \begin{equation}\label{Eq.:8}
\begin{split}
{\bf E}_{n}^{ag}=\hbar(G({\bf E}_{n-1}^{ag},{\bf W}_{n}^{ag})+{\bf E}_{n-1}^{ag}),
\end{split}
\end{equation}
where $\hbar$ is the $tanh$ activation function. $G$ means the residual mapping to be learned. ${\bf W}_{n}^{ag}$ indicates the to-be-learned parameter. Here,
the bias vector is absorbed into ${\bf W}_{n}^{ag}$ for simplicity.

Third, after three stacked 1D residual blocks, the output from the last 1D residual block is input into two additional FC layers for generating the final audio waveform ${\bf{a}}^{gen}$.

Finally, the generated audio waveform ${\bf{a}}^{gen}$ is leveraged to calculate the loss function for optimize the audio generation sub-network. Following the previous work \cite{chen2018visually}, a smooth L1 regression loss function is utilized to make the generated audio waveform as similar as the supervised audio. The loss function (called generation loss) is formulated as:
\begin{equation}\label{Eq.:9}
\begin{split}
\mathcal{L}_{gen}=\Gamma({\bf{a}}_{i}^{tr}-{\bf{a}}_{i}^{gen})+\lambda_{4}\sum_{l=1}^{L_{4}}\sum_{t=1}^{T_{4}}({\rm W}_{lt}^{gen})^{2},
\end{split}
\end{equation}
in which,
\begin{equation}\label{Eq.:10}
\begin{split}
\Gamma(x)=
\left\{
\begin{array}{lr}
0.5x^{2},&if\ \arrowvert x \arrowvert<1\\
\\
\arrowvert x \arrowvert-0.5, &othervise\\
\end{array}
\right.
\end{split}
\end{equation}
where subscript $i$ indexes the $i$-th training data. $x$ represents a real number, $\Arrowvert\cdot\Arrowvert_{1}$ indicates L1 norm. $\lambda_{4}$ is the coefficient of weight penalty which controls the relative importance of the two terms in Eq. \ref{Eq.:9}. ${\rm W}_{lt}^{gen}$ means the value in row $l$ and column $t$ in matrix ${\bf{W}}^{gen}$, which denotes all the to-be-learned parameters in the audio generation sub-network. $L_{4}$ and $T_{4}$ are the size dimensions of ${\bf{W}}^{gen}$.

Note that the autoencoder architecture in the audio feature learning part is not adopted for generating audio sample because of its limitation on modeling complex relationship of different phonemes. Throughout the audio generation sub-network, the raw waveform is obtained and interpolated to generate the corresponding audio file according to the sample frequency.

\subsection{Training stage}
As is shown in Figure \ref{fig:2}, the proposed MT-Net is composed of three progressive sub-networks: 1) feature learning, 2) cross-modal mapping, and 3) audio generation. First, the training image sample with minibatch in training image set $I^{tr}$ is fed into the image feature learning part for learning image semantic feature. The input training audio sample with minibatch in training audio set $A^{tr}$ is fed to into the audio feature learning part for learning audio semantic feature. Notably, the audio feature learning part is trained using a self-supervised manner. By minimizing the loss in Eq. \ref{Eq.:6}, the low-dimensional audio semantic feature set is obtained. The learning of the image feature learning part is supervised by the human perceived category according to Eq. \ref{Eq.:4}. Second, the image semantic feature is sent into the cross-modal mapping sub-network for mapping the visual modality feature to a cross-modal representation with the same semantic concept as the audio feature. The learning of the cross-modal mapping sub-network is supervised by the obtained audio semantic feature according to Eq. \ref{Eq.:7}. Finally, the obtained cross-modal representation from the cross-modal mapping sub-network is input into the audio generation sub-network for generating the audio waveform. The learning of the audio generation sub-network is supervised by the input audio waveform according to Eq. \ref{Eq.:9}. The RMSProp optimizer is leveraged to train the network.

\begin{algorithm}[t]
\renewcommand{\algorithmicrequire}{\textbf{Input:}}
\renewcommand\algorithmicensure {\textbf{Output:} }

\caption{Proposed MT-Net}
\label{alg:Framwork}
\begin{algorithmic}[1]

\REQUIRE ~~\\

Training image set $I^{tr}=\left[{\bf{I}}_{1}^{tr}, {\bf{I}}_{2}^{tr}, \cdots, {\bf{I}}_{N}^{tr}\right]$;\\
Training audio set $A^{tr}=\left[{\bf{a}}_{1}^{tr}, {\bf{a}}_{2}^{tr}, \cdots, {\bf{a}}_{N}^{tr}\right]$; \\
Human perceived category $Y=\left[{\bf{y}}_{1},{\bf{y}}_{2},\cdots, {\bf{y}}_{N}\right]$;\\
Testing image set $I^{te}=\left[{\bf{I}}_{1}^{te}, {\bf{I}}_{2}^{te}, \cdots, {\bf{I}}_{M}^{te}\right]$.

\ENSURE ~~\\
Testing audio set $A^{te}=\left[{\bf{a}}_{1}^{te}, {\bf{a}}_{2}^{te}, \cdots, {\bf{a}}_{M}^{te}\right]$;\\
All the optimised parameters ${\bf{W}}^{J}$ in the network except for the audio feature learning part.

\renewcommand{\algorithmicrequire}{\textbf{Initialization:}}
\REQUIRE ~~\\
All weights are randomly sampled by truncated\_normal distribution.\\
\renewcommand{\algorithmicrequire}{\textbf{Repeat:}{\,}{(for each iteration and until a fixed number of iterations)}}
\REQUIRE ~~\\

\STATE Compute the audio semantic feature $\varphi^{A}$ of each training audio sample ${\bf{a}}^{tr}$;

\STATE Update the parameters ${\bf{W}}^{A}$ according to Eq. \ref{Eq.:6}.

\renewcommand\algorithmicensure {\textbf{End}}
\ENSURE ~~\\
\STATE Generate the audio semantic feature set $\Phi^{A}=[\varphi^{A}_{1}, \varphi^{A}_{2}, \cdots,\varphi^{A}_{N}]$

\renewcommand{\algorithmicrequire}{\textbf{Repeat:}{\,}{(for each iteration and until a fixed number of iterations)}}
\REQUIRE ~~\\

\STATE Compute the image semantic feature $\varphi^{I}$ of each training image sample ${\bf{I}}^{tr}$;

\STATE Compute the cross-modal representation $\varphi$ using the cross-modal mapping sub-network by approximating $f(\varphi^{I})$ and $\varphi^{A}$;

\STATE Generate the audio waveform using the audio generation sub-network;

\STATE Update the parameters ${\bf{W}}^{J}$ according to Eq. \ref{Eq.:11}.

\renewcommand\algorithmicensure {\textbf{End}}
\ENSURE ~~\\

\STATE Generate the testing audio set $A^{te}=\left[{\bf{a}}_{1}^{te}, {\bf{a}}_{2}^{te}, \cdots, {\bf{a}}_{M}^{te}\right]$ corresponding to the testing image set $I^{te}$.
\renewcommand\algorithmicensure {\textbf{Return:}{\,}{$A^{te}, {\bf{W}}^{J}$}}
\ENSURE ~~\\
\end{algorithmic}
\end{algorithm}

Note that the parameters of the audio feature learning part are learned separately through minimizing the reconstruction loss according to Eq. \ref{Eq.:4}. In contrast, the parameters of image feature learning, cross-modal mapping and audio generation are learned through minimizing the joint loss. The joint loss is a linear weighted combination of the perceptual loss in Eq. \ref{Eq.:6}, the representation loss in Eq. \ref{Eq.:7} and the generation loss in Eq. \ref{Eq.:9}. The joint loss can be written as:
 \begin{equation}\label{Eq.:11}
\begin{split}
\mathcal{L}_{joint}=&\eta_{1}\mathcal{L}_{per}+\eta_{2}\mathcal{L}_{rep}+\eta_{3}\mathcal{L}_{gen}\\
=&\eta_{1}\left(-\frac{1}{N}\sum_{i=1}^{N}\left({\bf{y}}_{i}^{\mathrm{T}}log(\widehat{{\bf{y}}_{i}})\right)\right)+\eta_{2}\left(1-\sum_{i=1}^{N}\varphi^{A}_{i}\varphi_{i}/\sqrt{\sum_{i=1}^{N}(\varphi^{A}_{i})^{2}}\sqrt{\sum_{i=1}^{N}(\varphi_{i})^{2}}\right)\\
&+\eta_{3}\Arrowvert\Gamma({\bf{a}}_{i}^{tr}-{\bf{a}}_{i}^{gen})\Arrowvert_{1}+\eta_{4}\sum_{l=1}^{L}\sum_{t=1}^{T}({\rm W}_{lt}^{J})^{2},
\end{split}
\end{equation}
where subscript $i$ indexes the $i$-th training data. $\eta_{1}$, $\eta_{2}$, $\eta_{3}$ and $\eta_{4}$ are the trade-off coefficients which control the relative importance of the four terms in Eq. \ref{Eq.:11}. ${\rm W}_{lt}^{J}$ indicates the value in row $l$ and column $t$ in matrix ${\bf{W}}^{J}$, which denotes all the to-be-learned parameters in the network except for the audio feature learning part. $L$ and $T$ are the size dimensions of ${\bf{W}}^{J}$.

The main training procedure of the proposed network is shown in Algorithm \ref{alg:Framwork}.

\subsection{Testing stage}
Once the entire network is trained, it can be used for generating audio description given a testing image sample. The testing network is different from the training network in some aspects. Specifically, the audio feature learning part and the softmax layer in the image feature learning part are removed. For each testing image sample, it is preprocessed to a standard normalization according to Eq. \ref{Eq.:1} firstly. Then, the normalized image is input into the testing network for inferring the corresponding audio waveform based on the learned parameter ${\bf{W}}^{J}$. Finally, the obtained audio waveform is interpolated to generate the corresponding audio file according to the sample frequency.

\section{Experiment and results}\label{experiments}
 In this section, datasets, evaluation metrics and implementation details about the proposed network are elaborated. In addition, the experimental results are presented and analyzed.

\subsection{Datasets}
Being the first attempt to explore the audio description generation problem, three large-scale datasets with plenty of manual audio descriptions for this task are built based on the existing MNIST \footnote{http://yann.lecun.com/exdb/mnist/}, CIFAR10 \footnote{https://www.cs.toronto.edu/~kriz/cifar.html} and CIFAR100 \footnote{https://www.cs.toronto.edu/~kriz/cifar.html} datasets. The details of built datasets are as follows:

{\it MNIST Audio Description Dataset:} Similar to MNIST dataset, the built MNIST audio description  (MNIST AD)  dataset consists of 60000 training examples and 10000 testing examples. The corresponding audio descriptions are provided for each example with standard American pronunciation. The length of the audio description varies from 322 milliseconds to 823 milliseconds.

{\it CIFAR10 Audio Description Dataset:} The CIFAR10 audio description (CIFAR10 AD) dataset consists of 60000 pairs of $32\times32$ colour images and audio descriptions. The length of the audio description varies from 359 milliseconds to 1089 milliseconds. There are 50000 training pairs and 10000 testing pairs.

{\it CIFAR100 Audio Description Dataset:} Just like the CIFAR10 audio description dataset, the CIFAR100 audio description (CIFAR100 AD) dataset is also composed of 60000 pairs of $32\times32$ colour images and audio descriptions. The length of the audio description varies from 315 milliseconds to 1331 milliseconds. There are 50000 training pairs and 10000 testing pairs.

\subsection{Evaluation metrics}
Three metrics are adopted for evaluating the proposed network, including 2D correlation (Corr2D) between the actual and generated auditory spectrogram \cite{akbari2018lip2audspec}, standard Perceptual Evaluation of Speech Quality (PESQ) \cite{8902088}, and Short-Time Objective Intelligibility (STOI) \cite{8462040}.

Corr2D \cite{akbari2018lip2audspec} is used to measure the accuracy of the generated audio description in the frequency domain by calculating the 2D correlation between the actual and generated auditory spectrogram. Let $\bm S^{act}$ and $\bm S^{gen}$ denote each actual and generated auditory spectrogram, Corr2D is formulated as follows:
 \begin{equation}\label{Eq.:12}
\begin{split}
r = \frac{\sum\limits_{m}\sum\limits_{n}\left(\bm S^{gen}-\overline{S^{gen}}\right)\left(\bm S^{act}-\overline{S^{act}}\right)}{\sqrt{ \left(\sum\limits_{m}\sum\limits_{n}\left(\bm S^{gen}-\overline{S^{gen}}\right)^{2}\right)\left(\sum\limits_{m}\sum\limits_{n}\left(\bm S^{act}-\overline{S^{act}}\right)^{2}\right)}}
\end{split}
\end{equation}
where $\overline{S^{act}}$ and $\overline{S^{gen}}$ are the mean value of the actual and generated auditory spectrogram, and $m$ and $n$ index the element in $\bm S^{gen}$ and $\bm S^{act}$.

PESQ \cite{8902088} is a particularly developed methodology for modeling subjective tests commonly used in telecommunications to assess the voice quality by human beings. PESQ is implemented by comparing the actual audio signal with the generated signal passed through a communication system. After the PESQ evaluation, a score is given ranging from -0.5 to 4.5. A higher score means a better speech quality.

STOI \cite{8462040} is used to measure the intelligibility of the generated audio description. Specifically, STOI is a function of the actual and generated audio description, which are first decomposed into DFT-based, one-third octave bands. Next, short-time temporal envelope segments of the actual and generated audio description are compared by means of a correlation coefficient. Before comparison, the short-time generated audio description temporal envelopes are first normalized and clipped. These short-time intermediate intelligibility measures are then averaged to one scalar value, which is expected to have a monotonic increasing relation with the speech intelligibility. After the STOI evaluation, a score is given ranging from 0 to 1. A higher score means a better speech quality.

\subsection{Implementation details}
The proposed network includes three main sub-networks: 1) feature learning, 2) cross-modal mapping, and 3) audio generation. The feature learning sub-network is composed of image feature learning and audio feature learning. During the training phase, note that the audio feature learning part is trained separately, while image feature learning, cross-modal mapping and audio generation are trained jointly. The image feature learning part consists of 5 stacked residual blocks with 2 fully connected layers for encoding the image feature, which is summarized in Table \ref{tab:1}. The audio feature learning part consists of 6 stacked FC layers for encoding the audio feature, which is summarized in Table \ref{tab:2}. $L_{s}$ in Table \ref{tab:2} denotes the length of the input audio waveform ($L_{s}=18140$ for MNIST audio description dataset, $L_{s}=24020$ for CIFAR10 audio description dataset and $L_{s}=29356$ for CIFAR100 audio description dataset). Once the audio feature learning part is trained, the output of $\rm{FC}_{3}^{2}$ is considered as the learned condensed audio semantic feature. The learned condensed audio semantic feature is used as the supervised information for jointly training image feature learning, cross-modal mapping and audio generation sub-networks. The three sub-networks are trained using the joint loss in Eq. \ref{Eq.:11}, where the trade-off coefficients $\eta_{1}$, $\eta_{2}$, $\eta_{3}$ and $\eta_{4}$ in Eq. \ref{Eq.:11} are set as 0.5, 1.0, 1.0 and 0.8.

When training the MT-Net, the shorter audios of each dataset are zero-padded to the maximal temporal length of the audio signal in that dataset. Specifically, the audios in MNIST, CIFAR10 and CIFAR100 audio description datasets are adjusted to 823, 1089 and 1331 milliseconds, respectively. As for the training phase, the minibatch input is fed into the network, and reconstruction loss (Eq. \ref{Eq.:6}) and the joint loss (Eq. \ref{Eq.:11}) are minimized using RMSprop optimizers. The initial learning rate is set as 0.001. The weight decay is set as 0.0005. And the momentum is set as 0.9. The experiment is implemented with the $tensorflow$ library using a TITAN X (Pascal) GPU with 64G RAM for accelerating.
\begin{table}
\caption{The architecture of image feature learning part.}
\begin{center}
\begin{spacing}{1.19}
\small
\setlength{\tabcolsep}{0.9mm}{
\begin{tabular}{cccccccccccc} \toprule
Layers    & Trans1    & Res\_conv1x  &Trans2 & Res\_conv2x &Trans3 &Res\_conv3x \\ \hline
\specialrule{0em}{2pt}{2pt}
Parameters   & $1\times1, 32$  & $\begin{bmatrix} 3\times3,32 \\ 3\times3,32 \end{bmatrix}$ & $1\times1, 128$ & $\begin{bmatrix} 3\times3,128 \\ 3\times3,128 \end{bmatrix}$  & $1\times1, 256$ &$\begin{bmatrix} 3\times3,256 \\ 3\times3,256 \end{bmatrix}$ \\
\specialrule{0em}{2pt}{2pt}
\hline
Layers &Trans4 &Res\_conv4x &Trans5 &Res\_conv5x & $\rm{FC}_{1}^{1}$ & $\rm{FC}_{2}^{1}$\\ \hline
\specialrule{0em}{2pt}{2pt}
Parameters & $1\times1, 512$ &$\begin{bmatrix} 3\times3,512 \\ 3\times3,512 \end{bmatrix}\rule{0ex}{2.1em}$ & $1\times1, 1024$ &$\begin{bmatrix} 3\times3,1024 \\ 3\times3,1024 \end{bmatrix}$& 1024 & 1024\\
\specialrule{0em}{2pt}{2pt}
\bottomrule
\end{tabular}}
\end{spacing}
\end{center}
 \label{tab:1}%
\end{table}

\begin{table}
\caption{The architecture of audio feature learning part.}
\begin{center}
\begin{spacing}{1.29}
\begin{tabular}{cccccccc} \toprule
Layers    & $\rm{FC}_{1}^{2}$    & $\rm{FC}_{2}^{2}$  &$\rm{FC}_{3}^{2}$ & $\rm{FC}_{4}^{2}$ & $\rm{FC}_{5}^{2}$ & $\rm{FC}_{6}^{2}$ \\ \hline
In Size   & $L_{s}$  & 4096  & 2048  & 1024  & 2048 &4096 \\
Out Size & 4096  & 2048  & 1024  & 2048 & 4096 & $L_{s}$\\
\bottomrule
\end{tabular}
\end{spacing}
\end{center}
 \label{tab:2}%
\end{table}

\subsection{Ablation analysis}
In this subsection, several variations of the proposed MT-Net are constructed to examine: 1) the importance of representation loss function (see Meth.($\mathcal{L}_{gen}$) and Meth.($\mathcal{L}_{gen}+\mathcal{L}_{rep}$) networks); 2) the importance of perceptual loss function (see Meth.($\mathcal{L}_{gen}+\mathcal{L}_{rep}$) and the proposed networks); 3) the effect of the convolution kernel with holes (see proposed(hole) and the proposed networks); 4) the effect of the proposed audio generation sub-network (see proposed(autuencoder) and the proposed networks).

Specifically, the {\bf Meth.($\mathcal{L}_{gen}$)} only adopts the generation loss (Eq. \ref{Eq.:9}) for audio description generation. The {\bf Meth.($\mathcal{L}_{gen}+\mathcal{L}_{rep}$) } applies the combination of generation loss and representation loss (Eq. \ref{Eq.:11} but without the perceptual loss term)  for audio description generation. The {\bf proposed(hole)} network adopt the architecture of the proposed network but adopting the convolution kernel without holes. The {\bf proposed(autoencoder)} replaces the proposed audio generation sub-network with the autuencoder architecture in audio feature learning part.

\begin{table}
\caption{Quantitative evaluation of the proposed network for ablation analysis.}
\begin{center}
\begin{spacing}{1.19}
\small
\setlength{\tabcolsep}{0.6mm}{
\begin{tabular}{c|ccc|ccc|ccc} \toprule[1pt]
    \multirow{2}[4]{*}{Networks} & \multicolumn{3}{c|}{MNIST AD Dataset} & \multicolumn{3}{c|}{CIFAR10 AD Dataset} & \multicolumn{3}{c}{CIFAR100 AD Dataset} \\
\cline{2-10}          & Corr2D & PESQ  & STOI  & Corr2D & PESQ  & STOI  & Corr2D & PESQ  & STOI \\
\midrule[1pt]
    Meth.($\mathcal{L}_{gen}$)           &0.8560      & 2.3041       & 0.8925        & 0.6867        & 1.1896    &0.6051         & 0.5470    & 0.8912        & 0.3723 \\
    Meth.($\mathcal{L}_{gen}$+$\mathcal{L}_{rep}$)           &0.9147      & 2.9317       & 0.9521        & 0.7402        & 1.5172    & 0.6617        & 0.5742    & 1.1282        & 0.3941 \\
     \hline
     Proposed(hole)         & 0.9351     & 3.0424       & 0.9653        & 0.7528        & 1.5046    & 0.6739        & 0.5857    & 1.1846        & 0.4218 \\
    Proposed(autoencoder)          & 0.7491     & 1.8784       & 0.7962        & 0.5213        & 0.9257    & 0.5575        & 0.4812    & 0.7162        & 0.3384 \\
    \hline
    \bf MT-Net     & \bf0.9427 & \bf3.1577   & \bf0.9670    & \bf0.7654   & \bf1.7574 & \bf0.6996   & \bf0.6053 & \bf1.3050   & \bf0.4465 \\
\bottomrule[1pt]
\end{tabular}}
\end{spacing}
\end{center}
 \label{tab:3}%
\end{table}

The experiment result of quantitative evaluation for ablation analysis is reported in Table \ref{tab:3}.
First, by comparing the results of Meth.($\mathcal{L}_{gen}$) and Meth.($\mathcal{L}_{gen}+\mathcal{L}_{rep}$) networks, it can be found that the representation loss function is important to improve both quality and intelligibility of the generated audio description. In other words, learning a low-dimensional audio semantic feature for supervising the cross-modal mapping is quite important for generating better audio description.
 This may be due to two key facts: 1)  the raw waveform contains large variability in timbre, pitch and amplitude, which hampers the semantic information learning; 2) the raw waveform is quite long (tens of thousands of dimensions), which hinders the mapping of the semantic information from visual modality to audio modality when adopted as the supervised information directly.
 Second, by comparing the results of Meth.($\mathcal{L}_{gen}+\mathcal{L}_{rep}$) and the proposed networks, we can find the proposed network achieves more better performance, which  is because the proposed network captures more discriminative semantic information under the constraint of the perceptual loss.
 Third, according to the comparison results of proposed(hole) and the proposed networks, we can conclude that the convolution kernel with holes contributes to improving both quality and intelligibility of the generated audio description for enlarging the receptive field so as to capture more context information.
Finally, from the comparison results of proposed(autoencoder) and the proposed networks, the proposed network surpasses proposed(autoencoder) network to a great extent surprisingly, which demonstrates the proposed MT-Net is more effective than the autoencoder architecture in audio feature learning part.

\begin{table}
\caption{Quantitative evaluation of the proposed network compared to the state-of-the-arts.}
\begin{center}
\begin{spacing}{1.19}
\setlength{\tabcolsep}{1.0mm}{
\begin{tabular}{c|ccc|ccc|ccc} \toprule[1pt]
    \multirow{2}[4]{*}{Networks} & \multicolumn{3}{c|}{MNIST AD Dataset} & \multicolumn{3}{c|}{CIFAR10 AD Dataset} & \multicolumn{3}{c}{CIFAR100 AD Dataset} \\
\cline{2-10}    & Corr2D         & PESQ         & STOI       & Corr2D    & PESQ       & STOI      & Corr2D      & PESQ      & STOI \\
\midrule[1pt]
    DCMAVG      &0.8759          & 2.6305       & 0.9274     & 0.7168    & 1.3937     &0.6325     & 0.5526      & 1.0412    & 0.4073 \\
    CMCGAN      &0.9336          & 3.1627       & 0.9542     & 0.7406    & 1.5882     & 0.6737    & 0.5753      & 1.2036    & 0.4254 \\
    I2T2A       &0.9126          & 2.9614       & 0.9451     & 0.7263    & 1.6104     & 0.7122    & 0.5860      & 1.2642    & 0.4651 \\
    \bf MT-Net   &\bf0.9427      & \bf3.1577    & \bf0.9670  & \bf0.7654 & \bf1.7574  & \bf0.6996 & \bf0.6053   & \bf1.3050 & \bf0.4465 \\
\bottomrule[1pt]
\end{tabular}}
\end{spacing}
\end{center}
 \label{tab:4}%
\end{table}

\subsection{Comparison with state-of-the-arts}
Since the I2AD task is a totally new task proposed by us for the first time, there are currently no methods specifically for this task. To be convinced, we reproduced the codes of three existing methods which can be used to generate audio descriptions. The comparison methods include DCMAVG \cite{chen2017deep}, CMCGAN \cite{hao2018cmcgan} and I2T2A \cite{iccv_LiuT0G19} methods. Specifically, DCMAVG network is composed of a visual-to-audio GAN network and an audio-to-visual GAN network. The two GAN networks are trained separately. CMCGAN network performs the cross-modal visual-audio generation by building a cross-modal cycle GAN network, which unifies visual-to-audio and audio-to-visual into a common framework by a joint corresponding adversarial loss. The I2T2A method firstly generates the text description from image based on the image caption model  \cite{iccv_LiuT0G19}, and then transform the text description into corresponding audio based on the {\it Reading Woman} software \footnote{http://www.443w.com/tts/?sort=1}. The objective that we adopt I2T2A method for comparison is to demonstrate the advantage of the proposed I2AD scheme compared with the strategy of image-to-text and then text-to-audio. For fair comparison, the codes of these comparison methods are reproduced with their recommended parameter settings on MNIST, CIFAR10 and CIFAR100 AD datasets, respectively. Since the generated audio forms are spectrograms for the DCMAVG and CMCGAN networks, a three-convolution-layers with three-fully-connected-layers network is built to transform spectrogram into audio waveform. The comparison result is reported in Table \ref{tab:4}.

From Table \ref{tab:4}, it can be seen that the proposed MT-Net and CMCGAN network exceeds DCMAVG network in a large extent in all of the evaluation metrics on three datasets. In addition, the proposed MT-Net almost surpasses CMCGAN network, especially on more challenging CIFAR10 and CIFAR100 AD datasets. This may be due to the proposed MT-Net alleviating the semantic gap of intra-modal data and the heterogeneous gap of inter-modal data to a greater extent by the perceptual loss and representation loss. Compared with the I2T2A method, the proposed MT-Net performs better, especially on the MNIST AD dataset, which is because the proposed MT-Net generates audio descriptions from images straightly in an end-to-end manner and can preserve more implicit information. The STOI scores of the proposed MT-Net on CIFAR10 and CIFAR100 AD datasets cannot outperform the I2T2A method. This is because the I2T2A method adopts the classification manner for generating the text description. Although it can generate intelligible text description for challenging images, some implicit information, such as emotion and tone, will be lost due to the two-stage cross-modal conversion leading to more superposed loss. In contrast, the proposed MT-Net can preserve these information because the audio description is generated straightly in a learning manner. Overall, the proposed MT-Net is more superior than the existing state-of-the-arts.

\begin{figure}[tp]
\begin{center}
\includegraphics[width=1.0\linewidth]{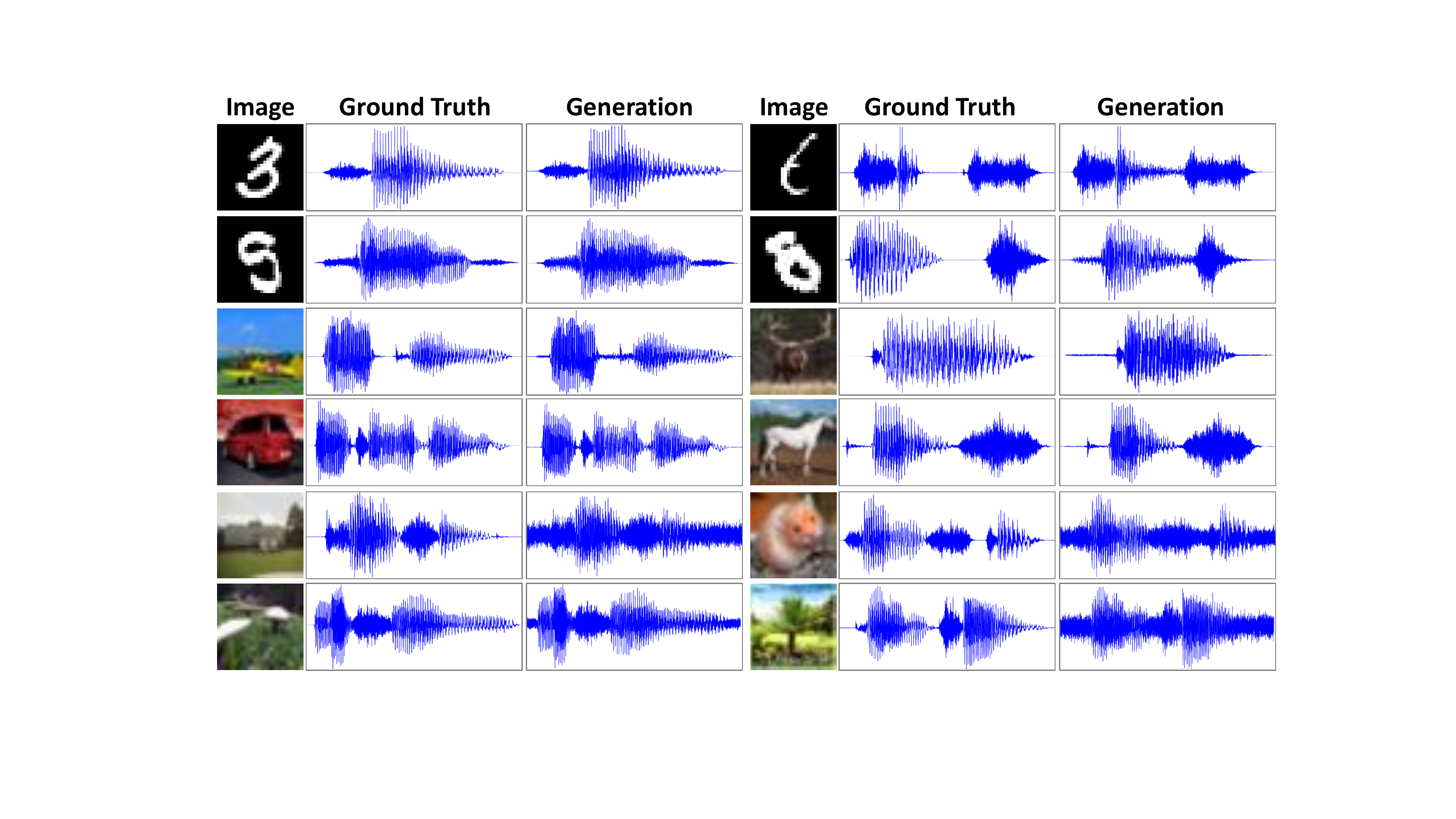}
\renewcommand{\figurename}{Fig.}
\end{center}
    \caption{\small{The qualitative results of the proposed MT-Net. The first two rows show some results on the MNIST AD dataset. The second row exhibit some results on the CIFAR10 AD dataset. The last two rows present some results on the CIFAR100 AD dataset.}}
\label{fig:5}
\end{figure}

\subsection{Qualitative results}
In order to illustrate the effectiveness of the proposed MT-Net more vividly, we visualized some examples of the generated audio descriptions on MNIST, CIFAR10 and CIFAR100 AD datasets, respectively. The qualitative results are shown in Figure \ref{fig:5}. As depicted in Figure \ref{fig:5}, the generated audio descriptions are very similar to the ground truth, especially on the MNIST AD dataset. It demonstrates the proposed MT-Net is indeed effective for generating audio descriptions from images. The generated audio descriptions on the CIFAR100 AD dataset are accompanied by some blurs and noises, because the task is more challenging to conduct the joint learning of semantic content and implicit information ({\it i.e.} emotion and tone).

\section{Conclusions}  \label{conclutions}
In this paper, an image-to-audio-description (I2AD) task is proposed to generate audio descriptions from images by exploring the inherent relationship of image and audio. To explore the task, a modal translation network (MT-Net) from visual to auditory sense is presented to generate audio descriptions from images. The proposed MT-Net can be used to assist visually impaired people to better perceive the environment. In order to evaluate the proposed network, three large-scale audio description datasets are built, {\it i.e.}, MNIST, CIFAR10 and CIFAR100 audio description datasets. Experiments on the built datasets demonstrate that the proposed network can indeed generate intelligible audio descriptions from visual images to a good extent. Moreover, the proposed network achieves the superior performance compared with other two similar state-of-the-art networks. In addition, some main components in the proposed network are proved to be effective. In the subsequent work, more complex form in the I2AD task will be explored to generate complete sentence in audio form by fully reducing the heterogeneous gap between image and audio modalities. We hope that visually impaired people will benefit from the exploration of the I2AD task eventually, allowing them to better perceive the world around them.

\section*{Acknowledgements}
This work was supported in part by the National Natural Science Found for Distinguished Young Scholars under Grant 61925112, in part by the National Natural Science Foundation of China under Grant 61806193, Grant 61702498, and Grant 61772510, in part by the Young Top-Notch Talent Program of Chinese Academy of Sciences under Grant QYZDB-SSW-JSC015, in part by the CAS ``Light of West China'' Program under Grant XAB2017B26, and Grant XAB2017B15, in part by the Natural Science Basic Research Program of Shaanxi under Grant 2019JQ-340, in part by the Key Research Program of Frontier Sciences, Chinese Academy of Sciences under Grant QYZDY-SSW-JSC044, in part by the National Natural Science Found for Distinguished Young Scholars under Grant 61825603, in part by the State Key Program of National Natural Science of China under Grant 61632018.

\section*{References}

\bibliography{mybibfile}

\end{document}